\documentclass[11pt]{article}
\usepackage{graphicx}
\usepackage{amsmath}
\usepackage{amssymb}
\usepackage{amsxtra}

\newcommand{\lsim}{\mathrel{\mathop{\kern 0pt \rlap
  {\raise.2ex\hbox{$<$}}}
  \lower.9ex\hbox{\kern-.190em $\sim$}}}
\newcommand{\gsim}{\mathrel{\mathop{\kern 0pt \rlap
  {\raise.2ex\hbox{$>$}}}
  \lower.9ex\hbox{\kern-.190em $\sim$}}}

\title{DAMA/LIBRA results and perspectives}

\author{R. Bernabei, P. Belli, S. d'Angelo, A. Di Marco, F. Montecchia\footnote{also Dip. di Ingegneria Civile e 
        Ingegneria Informatica, Univ. ``Tor Vergata'', I-00133 Rome, Italy}\\
        Dip. di Fisica, Univ. ``Tor Vergata''\\ and INFN-Roma ``Tor Vergata'', I-00133 Rome, Italy\\
        F. Cappella, A. d'Angelo, A. Incicchitti\\Dip. di Fisica, Univ. di Roma ``La Sapienza''\\ 
        and INFN-Roma, I-00185 Rome, Italy\\
        V. Caracciolo, S. Castellano, R. Cerulli\\Laboratori Nazionali del Gran Sasso, I.N.F.N., Assergi, Italy\\
        C.J. Dai, H.L. He, X.H. Ma, X.D. Sheng, R.G. Wang, Z.P. Ye\footnote{also University of Jing Gangshan, Jiangxi, China}\\
        IHEP, Chinese Academy, P.O. Box 918/3, Beijing 100039, China}

\begin{document}
\maketitle

\begin{abstract}
The DAMA/LIBRA experiment is composed by about 250 kg of highly radiopure NaI(Tl). It 
is in operation at the underground Gran Sasso National Laboratory of 
the INFN. The main aim of the experiment is to investigate the Dark Matter (DM) 
particles in the Galactic halo by exploiting the model independent DM annual modulation signature. 
The DAMA/LIBRA experiment
and the former DAMA/NaI (the first generation experiment having
an exposed mass of about 100 kg) have released 
results corresponding to a total exposure of 1.17 ton $\times$ yr
over 13 annual cycles; they have provided a model
independent evidence of the presence of DM particles in the galactic halo at 8.9 $\sigma$ C.L.. 
The results of a further annual cycle, concluding the DAMA/LIBRA--phase1, 
have been released after this Workshop and are not included here.
In the  fall 2010 an important upgrade of the experiment have been performed.
All the PMTs of the NaI(Tl) detectors have been replaced with new ones having 
higher quantum efficiency with the aim to decrease the software energy threshold
considered in the data analysis.  
The perspectives of the running DAMA/LIBRA--phase2 will be shortly summarized.
\end{abstract}

\section{The DAMA/LIBRA set-up}
The DAMA project develops and uses low background scintillators. 
It consists of the following experimental set-ups:
i) DAMA/NaI ($\simeq$ 100 kg of highly radiopure NaI(Tl)) that took data for 7 annual cycles 
and completed its data taking on July 2002 \cite{Nim98,allDM,Sist,epj06,RNC,allRare};
ii) DAMA/LXe, $\simeq$ 6.5 kg liquid Kr-free Xenon enriched either in $^{129}$Xe or
in $^{136}$Xe \cite{DAMALXe};
iii) DAMA/R\&D, a facility dedicated to tests on prototypes and to perform experiments 
developing and using various kinds of low background
crystal scintillators in order to investigate various rare processes \cite{DAMARD};
iv) DAMA/Ge, where sample measurements and 
measurements dedicated to the investigation of several rare processes are carried out as well as 
in the LNGS STELLA facility \cite{DAMAGE}; v) DAMA/CRYS, a new small set-up to test prototype detectors;
vi) the second generation DAMA/LIBRA set-up, $\simeq$ 250 kg
highly radiopure NaI(Tl)) \cite{perflibra,modlibra,modlibra2,review,papep,cncn,IPP,muons12,pmts}.
Many rare processes have been studied with these set-ups obtaining competitive results.

The main apparatus, DAMA/LIBRA, is investigating the 
presence of DM particles in the galactic halo 
by exploiting the model independent DM annual modulation signature.
In fact, as a consequence of its annual revolution around the Sun, 
which is moving in the Galaxy traveling with respect to the Local Standard 
of Rest towards the star Vega near
the constellation of Hercules, the Earth should be crossed by a larger 
flux of Dark Matter particles around $\sim$2 June 
(when the Earth orbital velocity is summed to the one of the
solar system with respect to the Galaxy) and by a smaller one 
around $\sim$2 December (when the two velocities are subtracted). 
This DM annual modulation signature 
is very distinctive since the effect induced by DM particles must 
simultaneously satisfy all the following requirements:
(1) the rate must contain a component modulated according to a cosine
function; 
(2) with one year period; 
(3) with a phase that peaks 
roughly around $\sim$ 2nd June;
(4) this modulation
must be present only in a well-defined low energy
range, where DM particles can induce signals; (5) it must
be present only in those events where just a single detector,
among all the available ones in the used set-up, actually
``fires'' ({\it single-hit} events), since the probability that DM 
particles
experience multiple interactions is negligible; (6) the
modulation amplitude in the region of maximal sensitivity
has to be $\lsim 7\%$ in case of usually adopted halo distributions,
but it may be significantly larger in case of some particular
scenarios such as e.g. those in refs. \cite{Wei01,Fre04}.
At present status of technology it is the only model independent signature 
available in 
direct Dark Matter investigation that can be effectively exploited.

\vspace{0.3cm}
The DAMA/LIBRA data released at time of this Workshop correspond to
six annual cycles for an exposure of 0.87 ton$\times$yr 
\cite{modlibra,modlibra2}.
Considering these data together with those previously collected by 
DAMA/NaI
over 7 annual cycles (0.29 ton$\times$yr), the total exposure collected
over 13 annual cycles is 1.17 ton$\times$yr; this
is orders of magnitude larger than the exposures typically collected in 
the field.

The DAMA/NaI set up and its performances are described in 
ref. \cite{Nim98,Sist,RNC,ijmd}, while
the DAMA/LIBRA set-up and its performances are described in ref. 
\cite{perflibra,modlibra2}.
The sensitive part of the DAMA/LIBRA set-up is made of 25
highly radiopure NaI(Tl) crystal scintillators placed in a 5-rows by 
5-columns matrix;
each crystal is coupled to two low background photomultipliers working in 
coincidence at single photoelectron level. 
The detectors are placed inside a sealed copper box flushed with HP 
nitrogen
and surrounded by a low background and massive shield made of  
Cu/Pb/Cd-foils/polyethylene/paraffin; moreover, about 1 m concrete (made 
from the Gran
Sasso rock material) almost fully surrounds (mostly outside
the barrack) this passive shield, acting as a further neutron
moderator. The installation has a 3-fold levels sealing system
which excludes the detectors from environmental air. The
whole installation is air-conditioned and the temperature is
continuously monitored and recorded. 
The detectors' responses range 
from 5.5 to 7.5 photoelectrons/keV. Energy calibrations with 
X-rays/$\gamma$ sources 
are regularly carried out down to few keV
in the same conditions as the production runs. A
software energy threshold of 2 keV is considered.
The experiment takes data up to the MeV scale and thus it is also 
sensitive 
to high energy signals. For all the details see ref. \cite{perflibra}.

\section{Short summary of the results}

\begin{figure}[!ht]
\begin{center}
\includegraphics[width=\textwidth]{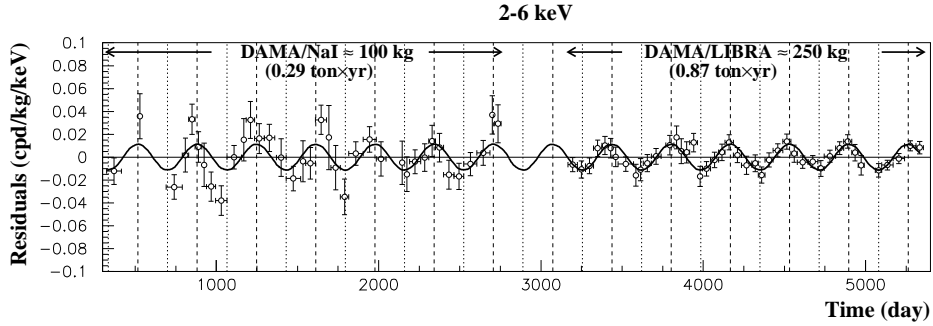}
\end{center}
\vspace{-0.8cm}
\caption{Experimental model-independent residual rate of the {\it 
single-hit} scintillation events,
measured by DAMA/NaI over seven and by DAMA/LIBRA over six annual cycles 
in the (2 -- 6) keV energy interval
as a function of the time \protect\cite{RNC,ijmd,modlibra,modlibra2}. The 
zero of the time scale is January 1$^{st}$
of the first year of data taking.
The experimental points present the errors as vertical bars and the 
associated time bin width as horizontal bars.
See refs.~\protect\cite{modlibra,modlibra2} and text.}
\label{fig1}
\end{figure}

Several analyses on the model-independent DM annual
modulation signature have been performed (see 
Refs.~\cite{modlibra,modlibra2,review} and references therein).
Here Fig. \ref{fig1} shows the time behaviour of the experimental
residual rates of the {\it single-hit}
events collected by DAMA/NaI and by DAMA/LIBRA in the (2--6) keV energy 
interval \cite{modlibra,modlibra2}.
The superimposed curve is the cosinusoidal function: $A \cos 
\omega(t-t_0)$
with a period $T = \frac{2\pi}{\omega} =  1$ yr, with a phase $t_0 = 
152.5$ day (June 2$^{nd}$),
and modulation amplitude, $A$, obtained by best fit over the 13 annual 
cycles.
The hypothesis of absence of modulation in the data can be discarded 
\cite{modlibra,modlibra2} and,
when the period and the phase are released in the fit, values well 
compatible
with those expected for a DM particle induced effect are obtained; for example,
in the cumulative (2--6) keV energy interval:
$A = (0.0116 \pm 0.0013)$ cpd/kg/keV, $T = (0.999 \pm 0.002)$ yr and $t_0 
= (146 \pm 7)$ day.
Summarizing, the analysis of the {\it single-hit} residual rate favours 
the presence of a
modulated cosine-like behaviour with proper features at 8.9$\sigma$ 
C.L.\cite{modlibra2}.

The same data of Fig. \ref{fig1} have also been investigated by a Fourier 
analysis obtaining 
a clear peak corresponding to a period of 1 year \cite{modlibra2}; 
this analysis in other energy regions shows instead only aliasing peaks.
It is worth noting 
that for this analysis the original formulas in 
Ref. \cite{sca0} have been slightly modified 
in order to take into account for the different time binning and the residuals errors
(see e.g. Ref. \cite{review}).

\begin{figure}[!ht]
\begin{center}
\includegraphics[width=\textwidth] {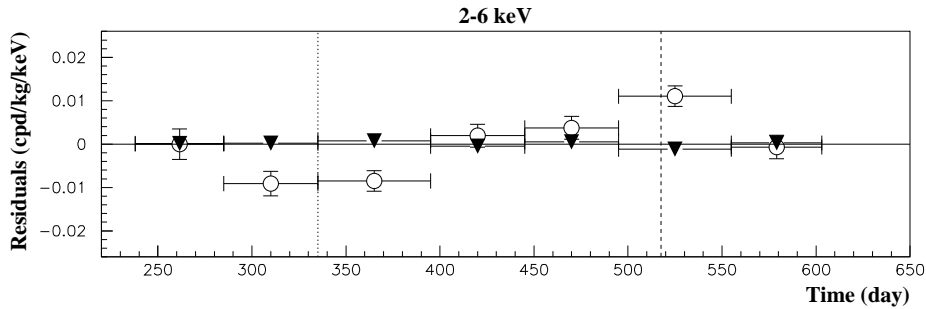}
\vspace{-0.8cm}
\caption{Experimental residual rates over the six DAMA/LIBRA annual cycles 
for {\it single-hit} events
(open circles) (class of events to which DM events belong) and for {\it 
multiple-hit} events (filled triangles)
(class of events to which DM events do not belong).
The initial time of the figure is taken on August 7$^{th}$.
The experimental points present the errors as vertical bars and the 
associated time bin width as horizontal
bars. See text and refs.~\protect\cite{modlibra,modlibra2}.
}
\label{fig_mul}
\end{center}
\end{figure}

Moreover, in order to verify absence of annual modulation in other energy regions and, thus,  
to also verify the absence of any significant background modulation, 
the time distribution in energy regions not of interest for DM detection 
has also been investigated: this allowed the exclusion of background
modulation in the whole energy spectrum at a level much
lower than the effect found in the lowest energy region for the {\it 
single-hit} events
\cite{modlibra2}. 
A further relevant investigation has been done by applying the same 
hardware and software 
procedures, used to acquire and to analyse the {\it single-hit} residual 
rate, to the 
{\it multiple-hits} events in which more than one detector ``fires''.
In fact, since the probability that a DM particle interacts in more than 
one detector 
is negligible, a DM signal can be present just in the {\it single-hit} 
residual rate.
Thus, this allows the study of the background behaviour in the same energy 
interval of the observed 
positive effect. The result of the analysis is reported in Fig. 
\ref{fig_mul} where 
it is shown the residual rate of the {\it single-hit} 
events measured over the six 
DAMA/LIBRA annual cycles, as collected in a single annual cycle, together 
with the residual rates of the {\it 
multiple-hits} events, in the same considered energy interval. 
A clear modulation is present in the {\it single-hit} events, while
the fitted modulation amplitudes for the {\it multiple-hits} 
residual rate are well compatible with zero \cite{modlibra2}.
Similar results were previously obtained also for the DAMA/NaI case 
\cite{ijmd}.
Thus, again evidence of annual modulation with proper features, as 
required by 
the DM annual modulation signature, is 
present in the {\it single-hit} residuals (events class to which the 
DM particle induced events belong), while it is absent in the {\it 
multiple-hits} residual rate (event class to 
which only background events belong).
Since the same identical hardware and the same identical software 
procedures have been used to analyze the 
two classes of events, the obtained result offers an additional strong 
support for the presence of a DM 
particle component in the galactic halo further excluding any side effect 
either from hardware or from software 
procedures or from background.

\begin{figure}[!ht]
\begin{center}
\includegraphics[width=\textwidth] {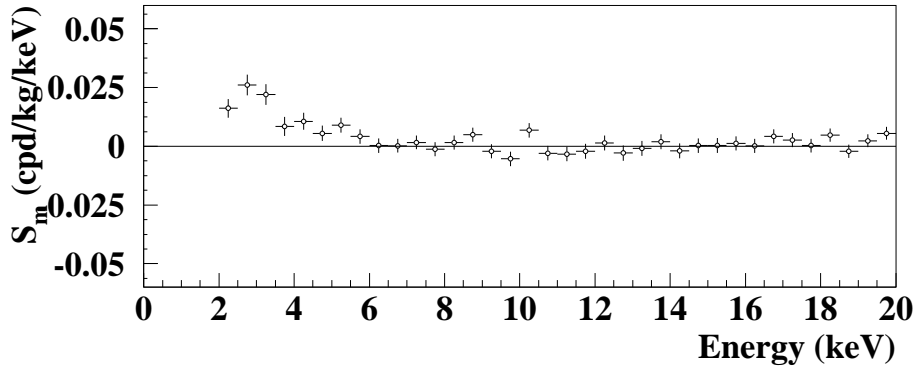}
\vspace{-0.8cm}
\caption{Energy distribution of the modulation amplitudes $S_{m}$ for the
total cumulative exposure 1.17 ton$\times$yr obtained by maximum 
likelihood analysis. 
The energy bin is 0.5 keV.
A clear modulation is present in the lowest energy region,
while $S_{m}$ values compatible with zero are present just above. 
See refs.~\protect\cite{modlibra,modlibra2} and text.}
\label{fig_sm}
\end{center}
\end{figure}

The annual modulation present at low energy has also been analyzed 
by depicting the differential modulation amplitudes, 
$S_{m}$, as a function of the energy; the $S_{m}$ is the
modulation amplitude of the modulated part of the signal obtained
by maximum likelihood method over the data, considering $T=1$ yr and 
$t_0=152.5$ day.
The $S_{m}$ values are reported as function of the energy in Fig. 
\ref{fig_sm}.
It can be inferred that a positive signal is present in the (2--6) keV 
energy interval, while $S_{m}$
values compatible with zero are present just above; in particular, the 
$S_{m}$ values
in the (6--20) keV energy interval have random fluctuations around zero 
with
$\chi^2$ equal to 27.5 for 28 degrees of freedom.
It has been also verified that the measured modulation amplitudes are 
statistically well
distributed in all the crystals, in all the annual cycles and energy bins;
these and other discussions can be found in ref. \cite{modlibra2}.

Many other analyses and discussions can be found in Refs. \cite{modlibra,modlibra2,review} 
and references  therein. 
Both the data of DAMA/LIBRA and of DAMA/NaI
fulfil all the requirements of the DM annual modulation signature.

Careful investigations
on absence of any significant systematics or side reaction have been quantitatively 
carried out (see e.g. Ref.
\cite{RNC,Sist,perflibra,modlibra,modlibra2,muons12,review,scineghe09,taupnoz,vulca010,canj11,tipp11,replica,replicaA}, 
and references therein). 
No systematics or side reactions able to mimic the signature (that is, able to
account for the measured modulation amplitude and simultaneously satisfy 
all the requirements of the signature) has been found or suggested
by anyone over more than a decade.

\begin{figure}[!ht]
\vspace{-.5cm}
\centering
\resizebox{0.45\columnwidth}{!}{%
\includegraphics{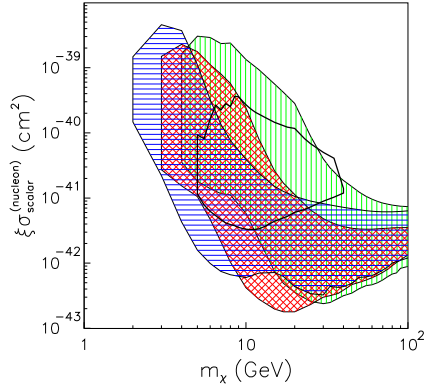} }
\vspace{-.5cm}
\caption{Regions in the nucleon cross section vs DM particle mass plane
allowed by DAMA for a DM candidate interacting via spin-independent
elastic scattering on target-nucleus; three different instances for
the Na and I quenching factors have been considered: i) without including
the channeling effect [(green) vertically-hatched region], ii) by including
the channeling effect [(blue) horizontally-hatched region)], and iii) without
the channeling effect considering energy-dependence of Na and I quenching
factors \cite{bot11} [(red) crosshatched region]. The velocity distributions and the
same uncertainties as in refs. \cite{RNC, ijmd} are considered here. These regions
represent the domain where the likelihood-function values differ more than
7.5 $\sigma$  from the null hypothesis (absence of modulation). The allowed
region obtained for the CoGeNT experiment, including the same astrophysical
models as in refs. \cite{RNC, ijmd} and assuming for simplicity a fixed value for the
Ge quenching factor and a Helm form factor with fixed parameters, is also reported
by a (black) thick solid line. This region includes configurations whose
likelihood-function values differ more than 1.64 $\sigma$  from the null hypothesis
(absence of modulation). For details see ref. \cite{bot11}.
}
\label{fig:reg}
\vspace{-0.1cm}
\end{figure}

The obtained DAMA model independent evidence
is compatible with a wide set of scenarios regarding the nature of the DM candidate
and related astrophysical, nuclear and particle Physics. For examples
some given scenarios and parameters are discussed e.g. in
Ref.~\cite{allDM,Sist,epj06,RNC,modlibra,review}.
Further large literature is available on the topics (see for example in Ref \cite{review}).
Moreover, both the negative results and all the possible positive hints, achieved so-far
in the field, are largely compatible with the DAMA model-independent DM annual
modulation results in many scenarios considering also the existing experimental and
theoretical uncertainties; the same holds for indirect approaches; see 
e.g. arguments in Ref. \cite{review} and references therein.
As an example in Fig. \ref{fig:reg} there are shown allowed regions for DM
candidates interacting by elastic scattering on target-nuclei with
spin-independent coupling, including also some of the existing uncertainties \cite{bot11}.

\section{DAMA/LIBRA--phase2 and perspectives}

A first upgrade of the DAMA/LIBRA set-up was performed in September 2008. 
One detector was recovered by replacing a broken PMT and a new optimization 
of some PMTs and HVs was done; the transient digitizers were replaced with new ones 
(the U1063A Acqiris 8-bit 1GS/s DC270 High-Speed cPCI Digitizers) having better 
performances and a new DAQ with optical read-out was installed. 
The DAMA/LIBRA--phase1 concluded its data taking in this configuration on 2010; 
the data of the last (seventh) annual cycle of this phase1 have been released 
after this Workshop \cite{libra-phase1}.

A further and more important upgrade has been performed at the end of 2010 when 
all the PMTs have been replaced with new ones having higher Quantum Efficiency (Q.E.), 
realized with a special dedicated development by HAMAMATSU co..
Details on the developments and on the reached performances in the operative conditions 
are reported in Ref. \cite{pmts}. 
We remind that up to October 2010 low background PMTs, developed by EMI-Electron Tubes with dedicated R\&D, 
were used; the light yield and other response features already 
allowed a software energy threshold of 2 keV in the data analysis.
The feasibility to decrease the software energy threshold below 2 keV in the new configuration
has been demonstrated\cite{pmts}. 

Since the fulfillment of this upgrade, the DAMA/LIBRA--phase2 is continuously running in order: 
(1) to increase the experimental sensitivity lowering the software energy threshold 
of the experiment; (2) to improve the corollary investigation on the nature of the 
DM particle and related astrophysical, nuclear and particle physics arguments; 
(3) to investigate other signal features. This requires long and 
heavy full time dedicated work for reliable collection and analysis of very large 
exposures, as DAMA collaboration has always done.

Another upgrade at the end of 2012 was successfully concluded:
new-concept preamplifiers were installed, with suitable operative and electronic features;
in particular, they allow the direct connection of the signal to the relative channel of the
Transient Digitizer (TD). 

Moreover, further improvements are planned. In particular, new trigger modules have been prepared and ready 
to be installed. A further simplification of the electronic chain has been proposed and funded; for such purpose
a new electronic module, New Linear FiFo (NLF), has been designed. It will allow -- among the others --
a significant reduction of the number of used NIM slots with definitive advantage.

In the future DAMA/LIBRA will also continue its study on several other rare
processes \cite{papep,cncn,IPP} as also the former DAMA/NaI apparatus did \cite{allRare}.

Finally, further improvements to increase the sensitivity of the set-up are under evaluation;
in particular, the use of high 
Q.E. and ultra-low background PMTs directly coupled to the
NaI(Tl) crystals is considered\footnote{However, this would require the 
disassembling of the detectors since the light guides act at present also as optical windows.}. 
This possible configuration will allow a further large improvement 
in the light collection and a further lowering of the software
energy threshold. Moreover, efforts towards a possible highly
radiopure NaI(Tl) ``general purpose'' experiment (DAMA/1ton)
having full sensitive mass of 1 ton (we already proposed in 1996
as general purpose set-up) are continuing in various aspects.


\end{document}